# Controlled photophoretic levitation of nanostructured thin films for near-space flight

**Authors:** Mohsen Azadi[1], George A. Popov[2], Zhipeng Lu[3], Andy G. Eskenazi[2], Ji Won Bang[1], Dr. Matthew F. Campbell[1], Prof. Howard Hu[1], and Prof. Igor Bargatin[1]

**Affiliations:** [1]Department of Mechanical Engineering and Applied Mechanics, University of Pennsylvania, Philadelphia, PA. [2]Vagelos Integrated Program in Energy Research, University of Pennsylvania, Philadelphia, PA. [3] Department of Chemistry, University of Pennsylvania, Philadelphia, PA.

*Correspondence and requests for materials should be addressed to bargatin@seas.upenn.edu.

**Abstract:** We report light-driven levitation of macroscopic polymer films whose bottom surface is engineered to maximize the thermal accommodation coefficient. Specifically, we levitated centimeter-scale disks made of commercial 0.5-micron-thick mylar film coated with carbon nanotubes on one side. When illuminated with light intensity comparable to natural sunlight, the polymer disk heats up and interacts with incident gas molecules differently on the top and bottom sides, producing a net recoil force. This lift force is maximized at gas pressures corresponding to Knudsen number on the order of 0.3, and correspondingly, we observed the levitation of 0.6-cm-diameter disks in a vacuum chamber at pressures between 10 and 30 Pa. Moreover, we controlled the flight of the disks using a shaped beam that optically trapped the levitating disks. Our experimentally validated theoretical model predicts that the lift forces can be many times the weight of the films, allowing payloads of up to 10 milligrams for sunlight-powered low-cost microflyers in the upper atmosphere at altitudes of 50-100 km.

Currently known flight mechanism cannot be used to achieve sustained flight over a long period of time in Earth's mesosphere, the upper layer of the atmosphere located at altitudes between ~50 and ~80 km[1]. Modern aircraft are not able to fly for an extended period of time above ~30-50 km because the air density at these altitudes is too low to generate lift for airplanes and balloons[2-4]. On the other hand, space satellites rarely dip below ~150 km because the air at such altitudes becomes thick enough to cause excessive drag and heating [5,6]. The only vehicles capable of flying in the mesosphere are rockets, which cannot be used for sustained flights.

Photophoresis or light-driven motion[7-9] can provide an alternative propulsion mechanism in the mesosphere[1,10]. Most recent photophoresis studies mostly focused on microscopic particles in atmospheric aerosols[11-15]. In the free molecular regime, when the mean free path $\lambda$ is much larger than the characteristic size $a$ of the object, the photophoretic force results from the difference in the velocity of the incident and departing gas molecules from a hot surface[16-19]. In contrast, in continuum regime ($\lambda \ll a$), the force is generated through the thermal creep of the gas over the edges of the sample from the cold side to the hot side[20-23]. The photophoretic force has been shown to reach a maximum in the transient regime, where the Knudsen number $Kn = \lambda/a$ is of order unity[24]. In this regime, a mix of free molecular and continuum mechanisms contribute to force generation[25], but even at its maximum, the typical value of photophoretic force for a centimeter-sized object is in the microNewton range[26]. Such low forces mean that the mass of the object needs to be in the milligram range or less to achieve levitation and that the needed temperature difference must be generated across a small thickness of an ultralightweight structure. We recently reported plate mechanical metamaterials that were designed to minimize their mass, generate a few degrees of temperature difference between its top and bottom, and to maximize the thermal creep using microchannels, which was sufficient to levitate such highly engineered structures[10,27].

However, it is also possible to generate a photophoretic force in ultrathin structures that have a negligible temperature difference but instead have different surface properties on the top and bottom. In the free-molecular regime, gas molecules colliding with a heated structure absorb energy from the surface and leave with a higher temperature. The measure for such energy transfer through gas-surface collisions is called the thermal accommodation coefficient, $\alpha = \frac{T_r - T_i}{T_s - T_i}$ [28,29]. Here, $T_r$ is the temperature of reflected gas molecules, $T_i$ and $T_s$ are incident molecule and structure temperature, respectively. For every combination of a surface material and gas species, there is a unique α, which depends on a variety of factors such as temperature, surface roughness, density and atomic/molecular weight of the surface and the gas, and even electronic properties of the surface[30-34]. If the thermal accommodation coefficient is larger on the bottom surface of a film, the momentum change of the gas molecules and the corresponding recoil of the structure is larger on the bottom side, resulting in a net lift force (Fig. 1A). This type of the photophoretic force is generated even if the top and bottom are at the same temperature, as long as these temperatures are higher than that of the ambient gas.

To demonstrate this approach, we fabricated macroscopic samples with submicron thickness and different surfaces on the top and bottom. By coating a mylar film with carbon nanotubes on only one side (Fig. 1B), we increased the thermal accommodation coefficient and generated a photophoretic force that levitated flat disks with centimeter-scale diameters. We showed that these levitating samples can be made using simple fabrication methods from low-cost materials and achieve stable mid-air hovering at pressures corresponding to altitudes of ~80 km in the atmosphere.

We used 500-nm-thick mylar film (also known as OS film in the model airplane community) and deposited a 300-nm-thick layer of carbon nanotubes (CNTs) on its bottom side (see supplementary information for detail). This layer acted as a lightweight light absorber with absorptivity of ~ 90% and also improved the sample's structural rigidity. The areal density of the resulting sample was ~ 1 gram per square meter with an overall thickness of ~ 0.8 μm. When illuminated with LED arrays, the structure became up to ~100 K hotter than the environment.

The CNT layer also has a nanostructured surface shown in the inset of Fig. 1B, which tends to trap incoming gas molecules as illustrated in Fig. 1A. These traps make the air molecules collide with the surface multiple times on average before leaving, resulting in a higher thermal accommodation coefficient for the CNT-air side compared to mylar-air side. This difference results in a higher departing velocity for the air molecules on the CNT side compared to the mylar side. The net momentum transfer from these gas-surface interactions results in an upward recoil force that levitates the sample, as shown in Fig. 1C. We note that this effect is not due to the temperature difference between the top and bottom, as in our previous experiments[10]. Using the thermal conductivity of mylar $k_{mylar} = 0.14 \frac{W}{mK}$ and air $k_{air} = 0.025 \frac{W}{mK}$ (which is the lower bound for the conductivity of the porous CNT layer), we estimate the temperature difference is less than $\Delta T_{max} = \frac{I_{light} t_{film}}{2 k_{air}}$ =0.1 K, where $I_{light} = 0.5 \frac{W}{cm^2}$ is the typical incident light intensity in our experiments, and $t_{film} \approx 1\ \mu m$ is the total thickness of the film and the CNT layer. Since such small temperature differences are insufficient to levitate the film using temperature-driven forces on disks or plates[10,35], the observed photophoretic force is not due to the temperature gradient and is instead a result of the difference in the accommodation coefficient of the two sides.

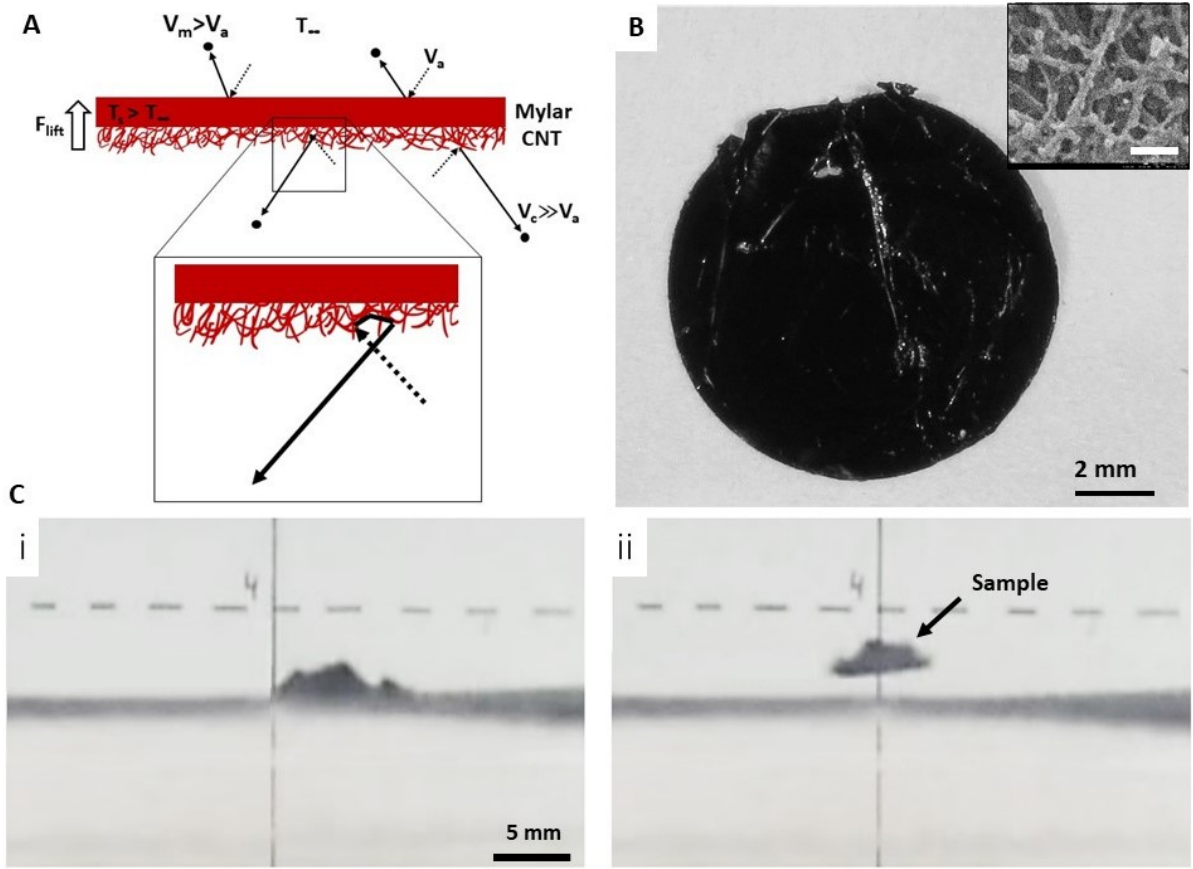

**Figure 1. A)** Schematic diagram of the main mechanism behind the photophoretic force due to a difference in the thermal accommodation coefficient (in the free molecular regime). **B)** Photograph of a 6-mm-diameter mylar disk covered by a layer of carbon nanotubes. The inset shows the porous surface of the CNT layer which traps incoming air molecules, allowing for the gas molecules to absorb more heat and approach unity thermal accommodation coefficient (scale bar 50 nm). **C)** Sequential screenshots of a levitating 6-mm-diameter disk under incident light intensity of $0.5 \frac{W}{cm^2}$.

To study the force generated by a difference in the thermal accommodation coefficient (Δα-force), we developed a theoretical model and calculated the areal density of an object that can be levitated under a certain flux and a known Δα. Briefly, in the free molecular regime, Δα-force increases proportionally with pressure and reaches a maximum at Knudsen numbers of order unity. Further increases in pressure reduce Δα-force as $P^{-2}$, which is faster than the $P^{-1}$ scaling that is typical for the temperature-difference forces (see supplementary information for full formulation).

Figure 2A shows the predicted areal density of an object that can be levitated using Δα-force with Δα=0.15±0.05 and flux of ~ 0.5 $\frac{W}{cm^2}$ (~5 times the direct sunlight intensity on the surface of the Earth and ~4 times the direct sunlight intensity in the upper atmosphere) as well as the results of our experiments with CNT-covered mylar disk. This value of Δα = 0.15±0.05 was found by fitting the theoretical predictions of successful levitation to experimental results (see supplementary Fig. S7). Figure 2B compares the pressure-dependent lift force to the weight of the 0.6-cm-diamater sample, with upward and downward arrows indicating levitation and no levitation, respectively, in experiments. We note that our mylar samples had an operational range limited by the maximum temperature they could sustain before thermal deformation. In particular, we observed the disks curl up at temepratures of ~400 K (see supplementary video 5 and Fig. S3 and S4). This temperature threshold was then used to map the operational range in Fig. 2, which matches the experiments well.

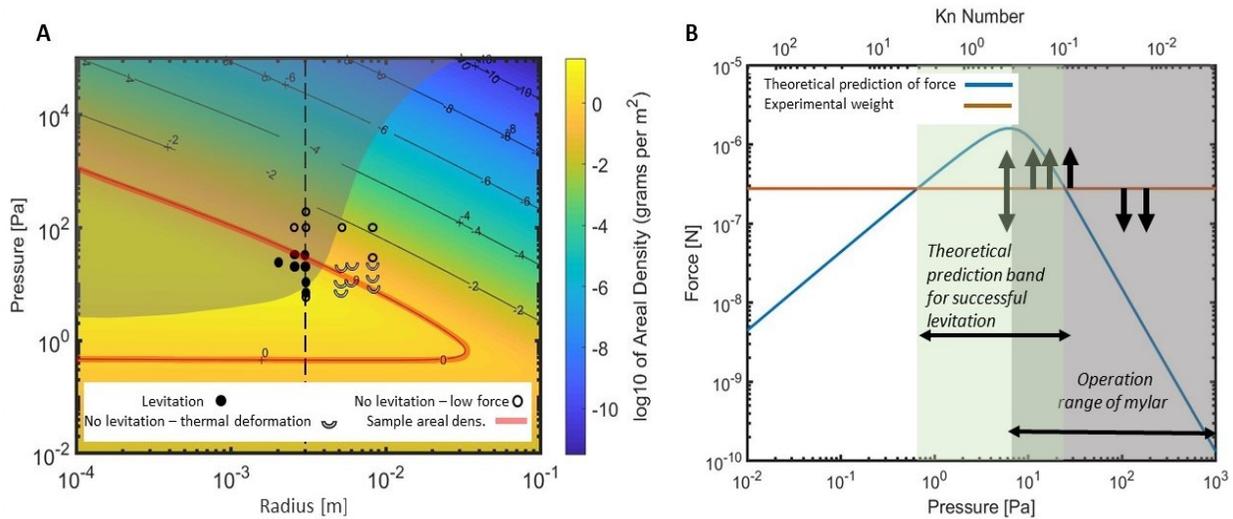

**Figure 2. A)** Areal density of an object with micron thickness that can be levitated under 0.5 $\frac{W}{cm^2}$ and $\Delta\alpha =$ 0.15. The shaded area represents the domain that mylar can operate without undergoing thermal deformation due to temperatures above 400 K (see supplementary information) **B)** comparison of the force and weight for a disk with 6-mm-diameter with thermal deformation considerations (the size corresponding to the dashed line in Fig. 3A).

In order to levitate samples for extended periods of time, we designed a light field that can optically trap the sample. This light trap consisted of a central area with intensity high enough to levitate the disk, surrounded by a ring of even higher intensity, which creates a restoring force by tilting the disk and pushing back toward the center (Fig. 3). Considering the dimensions and thermal properties of the sample material the thermalization time constant can be estimated as $\tau = \frac{\rho_{Mylar} C_{Mylar} V_{Mylar}}{h_{total} A_{disk}} = \frac{\rho_{Mylar} C_{Mylar} t_{Mylar}}{2 h_{total}} \sim 0.025\ s$, where $\rho_{Mylar} = 1390\ \frac{kg}{m^3}$ and $C_{mylar} = 1170\ \frac{J}{kg\ K}$ are the density and heat capacity of the mylar film, $A_{disk} = 2\pi a^2$ is the total area and $h_{total} = h_{cond} + h_{conv} + h_{rad}$ is the average heat transfer coefficient from the disk to the ambient and found numerically (see supplementary information) which for the successful experiments shown in Fig. 3A is $h_{total} \sim 17\ \frac{W}{m^2 K}$. When the incident light is kept at a constant intensity, the sample reaches thermal equilibrium after a few time constants, or about 0.05 second. Therefore, the light beam needs to be wide enough that the thermalization and the restoring force occur before the disk can cross the high-intensity ring, escaping from the trap.

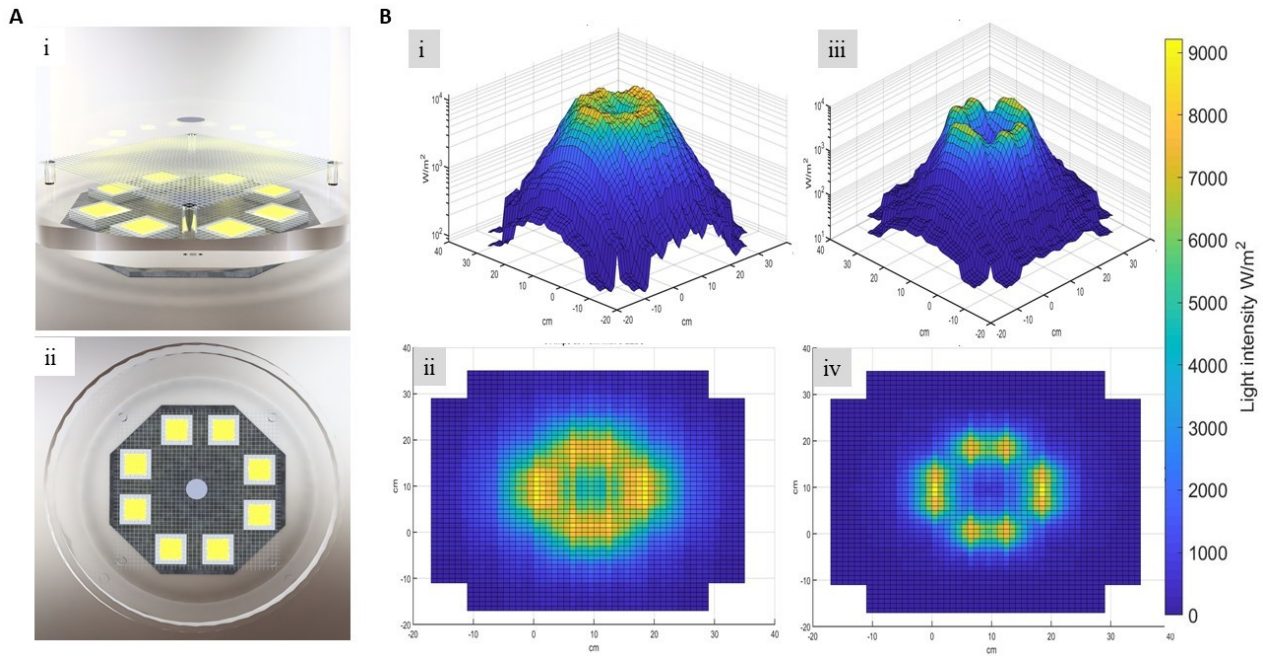

**Figure 3. A) i)** Side and **ii)** top-view schematic diagram of the test setup consisting of eight LED arrays below an acrylic vacuum chamber, a 74%-transparent metallic mesh placed several centimeters above the bottom surface of the acrylic chamber, and levitating disk sample. **B)** Experimental measurements of intensity of the trapping light beam from eight LEDs arrays at **i, ii)** 7-cm and **iii, vi)** 10-cm heights above the LEDs. Note that the high-intensity ring surrounding the microflyer confines its in-plane movement and that the intensity at the center drops as the height increases, which stabilizes the flight height.

We used two light traps to study this effect. The first consisted of four high-intensity arrays of LEDs placed in a square pattern, creating a light ring ~4 cm in diameter, which proved insufficient since the initial liftoff speed would push the samples out of the trap very quickly, before thermalization (see supplementary video 1). The second setup had eight high-intensity arrays of LEDs arranged in a ring pattern with a diameter of about 15 cm shown in Fig. 3B. In all experiments, a metallic mesh with a transparency of 74% was used as a launchpad and was placed several centimeters away from any inner surface of the vacuum chamber to avoid ground or wall effects. We also observed the samples levitate when the light was shone from above, demonstrating that this levitation mechanism can work for any direction of the incident light because the mylar film is optically transparent. However, creating a trapping beam configuration is more complex when illuminating from above and we did not pursue it.

Several tests using side and oblique video-recording revealed that the light trap was effective, and the samples levitated at a height comparable to their diameter above the mesh (see supplementary videos 2, 3, and 4). After several seconds of successful levitation, we typically increased the light intensity at a reate of $\sim 3 \frac{\text{kW}}{\text{m}^2}$ per minute, resulting in a gradual temperature increase that slowly deformed the sample after approximately 30 seconds. Once deformed, the lift force was reduced and a random side force appeared, occasionally pushing it outside of the light trap. In most cases, however, the deformation resulted in a lower effective light absorbing area, which then caused the sample to settle down within the light trap. At lower intensities, we expect the plates to remain in the light trap indefinitely.

Using our model, we can predict the possibility of photophoretic flight at different altitudes in the atmosphere. As the altitude increases from 0 to 100 km, ambient temperature and pressure change dramatically (see supplementary information and Fig. S6), which affects the temperature difference between the disk and ambient. In our model, we also accounted for different radiation environments seen by the top and bottom of the disk at altitudes above 30 km. Conservatively, we assumed a 3 K effective temperature for deep space, seen by the top side of the disk, and 255 K for the Earth, seen by the bottom side[36].

As shown in Fig. 4, our theoretical model predicts the possibility of sunlight-powered levitation in a wide range of altitudes between 50-100 km if the accommodation coefficient of the surfaces can reach $\Delta\alpha = 0.5$ and the thermal infrared emissivity is reduced to $\epsilon = 0.5$. Moreover, the disks can lift up to 10 mg of payload under natural sunlight (0.136 $\frac{\text{W}}{\text{cm}^2}$ in the upper atmosphere). Thermal accommodation coefficient value for air on clean glass and air on glass coated with molecularly thin polymer is reported to be 0.19 and 0.43, respectively[37], which means Δα=0.5 is realistic with carefully treated surfaces. The use of selective absorbers has been shown to reduce the emissivity to as low as $\epsilon \sim 0.1$[38], which would allow levitation even for incident light intensities below full natural sunlight intensity (see Fig. S11). Due to low ambient temperatures at high altitudes, the disk temperature can remain below 400 K (Fig. 4c), allowing the use of mylar or other polymer materials without thermal deformation.

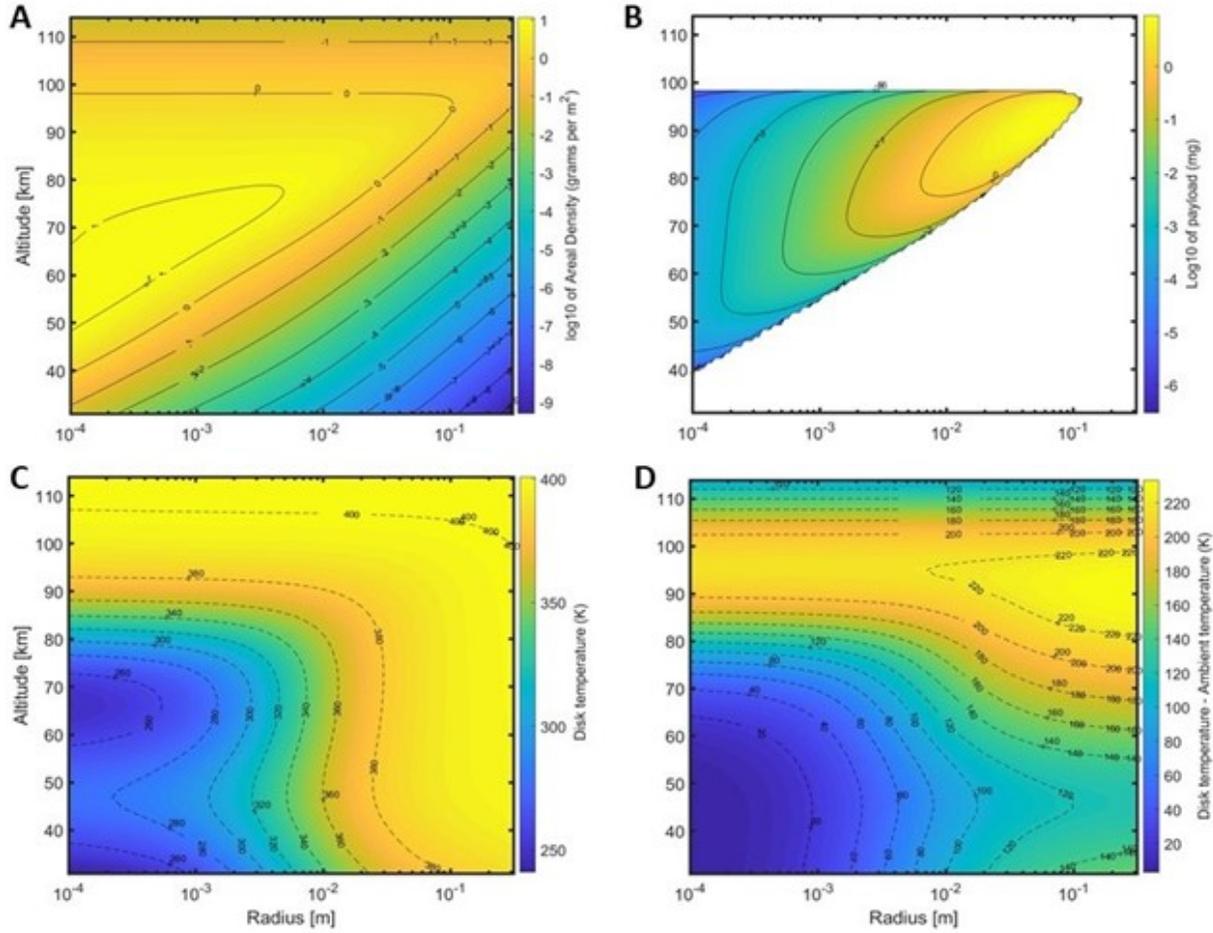

**Figure 4.** Contour plots of **A)** Areal density of the object able to be levitated **B)** Payload that can be lifted using mylar-CNT **C)** Temperature and **D)** Temperature difference between the disk and ambient for different sizes at different altitudes with $\Delta\alpha = 0.5$, $\epsilon = 0.5$ and under natural sunlight ($0.136 \frac{W}{cm^2}$).

In summary, this work demonstrated a new approach to photophoretic levitation of macroscopic structures that does not require a temperature gradient within the object, offering a path to the development of affordable photophoretic microflyers for the mesosphere. We developed a theoretical model for thin disks, which showed agreement with the experiments done using low-cost fabrication methods. The levitation tests were successful at pressures of ~10 Pa and incident light intensity of $0.5 \frac{W}{cm^2}$. We also presented a method to trap and control the hovering of the thin microflyers. Finally, photophoretic levitation through $\Delta\alpha$-force showed consistent flight direction regardless of the changes in the direction of incoming light.

Our experimentally validated model predicts that the same approach can be used in the near space at altitudes between 50 and 100 km. Such microflyers can use sunlight or a laser beam from any direction to stay levitated for extended periods of time, allowing, for example, the mapping of wind flows at these high altitudes by tracking the location of these flyers using a radar or lidar. There is a significant opportunity to further increase the force by increasing the difference in accommodation coefficients and reducing the infrared emissivity. Such improvements will allow the microflyers to carry payloads of up to 10 milligram, which can consist of thin-substrate sensors for weather and climate applications, such as measuring temperature, pressure, or carbon dioxide levels.

# Supplementary Information for

**Controlled photophoretic levitation of nanostructured thin films for near-space flight**

*Mohsen Azadi, George A. Popov, Zhipeng Lu, Andy G. Eskenazi, Ji Won Bang, Dr. Matthew F. Campbell, Prof. Howard Hu, and Prof. Igor Bargatin*

**This PDF file includes:**

Supplementary text
Figures S1 to S14
Captions for Movies S1 to S5

**Other Supplementary materials for this manuscript include the following:**

Movies S1 to S5

**Supplementary Text**

**Methods:**

Sample fabrication process

We started with a thin sheet of commercially available mylar film with a nominal thickness of 0.5 micron (Dupont). Using a $1 \text{cm}^2$ sample and a precision scale (Perkin-Elmer AD4 model), we measured the areal density to be ~ $0.7 \frac{\text{g}}{\text{m}^2}$, which agrees with the theoretical value expected from the nominal density of mylar of $1.39 \frac{\text{kg}}{\text{m}^3}$. To deposit the carbon nanotube (CNT) layer, we used a 0.2% w.t. water-based single-wall CNT with 1-2 nm diameter and 5-30 μm length (NanoAmor) and diluted it with DI water by a volumetric ratio of 3:1 (DI Water:CNT). We then stretched a sheet of this mylar thin film of a Si wafer and put it on a hot plate at 50 °C. By dropcasting the CNT solution on the sheet and letting the water evaporate, we created a CNT layer on the mylar sheet, then peeled the Mylar sheet off of the Si substrate and cut circular samples of the desired diameter using a razor blade. Weight measurements of the CNT-covered samples showed their areal density to be ~ 1 g/m².

Testing methods

The experimental setup used a 10-l custom-designed cylindrical acrylic vacuum chamber. The acrylic allowed for easy illumination of the sample from any direction and allows for video capture from any direction. Despite the 1-in thick walls and properly sealed junctions, the chamber leaked a significant amount of air through its walls (a known downside of acrylic chambers), making it impossible to reach high-vacuum base pressures. A two-stage vacuum pump with a 1500 Hz turbo pump resulted in base pressures ranging from 7 to 200 Pa (~0.05-1.5 Torr) by using only the roughing pump or roughing-turbo combination.

To create a light trap that has a local minimum in the center and a ring of maximum intensity, we used eight LED arrays, each rated for 100 W of input power (LOHAS LH-XP-100W-6000k). These LEDs, as shown in Figure S1, were mounted on two pieces of aluminum connected to 4 heat sinks with forced convection cooling from 4 fans, capable of removing ~1000 W in total. All thermal interfaces were enhanced using silver paste (Arctic Silver 5 Polysynthetic Thermal Compound). A stainless-steel mesh with a transparency of 74% was used as a "launchpad" (McMaster item # 9238T51), which we placed 3 cm above the bottom surface of the chamber to eliminate ground effects. Figure S1 shows a side view of the chamber, the eight-LED-array assembly, and the launchpad. To study the residual ground effect of the launchpad, we also tested an 85% transparent mesh which has only half as much covered area as the 74% mesh. The experimental results showed no measurable difference in the height vs input power to the LEDs, suggesting the effect of the mesh is negligible.

Theoretical Model:

Theoretical development for mid-air levitation of the structures requires in-depth understanding of heat transfer between the structure and environment. In this model, we start with the heat transfer analysis for a disk for the entire range of pressure. Then, we use the temperature distribution of the surface to find the temperature of the gas molecules impinging on and reflecting from the surface. Finally, we find the total amount of force experienced by a disk with two different surface properties on either side using a semi-empirical approach presented by Rohatschek[1].

*Heat Transfer Model:*

Force generation in free molecular and continuum regime obey distinct physics. Hence, the theoretical model, including the heat transfer model, should properly describe the physical phenomena in both regimes. We start by considering the energy balance for a disk and derive the equations for the surface temperature of the disk. In our model, the disk is absorbing radiation on one side and dissipating heat on both sides *via* radiation, convection, and conduction. In this approach, air is considered an ideal gas with the properties listed below:

heat capacity at constant pressure[2]

$$C_p \left[\frac{kJ}{kg\,K}\right] = 28.11 + (0.1956 \times 10^{-2})T[K] + (0.4802 \times 10^{-5})T[K]^2 - (1.966 \times 10^{-9})T[K]^3, \quad (1a)$$

thermal conductivity[3]

$$k_{air}\left[\frac{W}{m\,K}\right] = (0.238 \times 10^{-3})\,T[K]^{0.8218}, \quad (1b)$$

thermal diffusivity ($D$), thermal expansion coefficient ($\beta$), dynamic viscosity ($\mu$), and density given, respectively, by

$$D = \frac{k_{air}}{\rho c_p},\ \beta = \frac{1}{T_0},\ \mu = \mu_o \left(\frac{T}{T_0}\right)^{2/3},\ \text{and}\ \rho = \frac{P}{R_{air}T}, \quad (1c)$$

where $T$ is the temperature, $T_0 = 273$ K is the reference temperature, $\mu_o = 1.716 \times 10^{-5}$ Pa·s is dynamic viscosity at the reference temperature, $P$ is the pressure, and $R_{air} = \frac{R_u}{M_{air}} = 287.1 \frac{J}{kg\,K}$ is the ideal gas constant for air, obtained from the universal gas constant $R_u = 8.314 \frac{J}{mol\,K}$ and the molar mass of air $M_{air} = 0.02896 \frac{kg}{mol}$.

The incident energy is absorbed on one side of the disk and is balanced by the total heat transfer from the disk, which includes radiation, conduction, and convection, or

$$Q_{inc} = Q_{rad} + Q_{cond} + Q_{conv}. \quad (2)$$

Here $Q_{inc} = \frac{I_{inc}S}{2}$, $I_{inc}$ is the incident flux shone on one side of the disk and $S = 2\pi a^2$ is the total surface area of the disk. For simplicity, we assume the disk has a uniform temperature, $T_s$, which we found to be a reasonable approximation by comparing to the results of finite-element simulations in COMSOL under a variety of conditions. The radiative heat transfer from both sides of the disk is then given by

$$Q_{rad} = S\sigma\varepsilon(T_s^4 - T_\infty^4), \quad (3)$$

where $\sigma = 5.67 \times 10^{-8} \frac{W}{m^2 K^4}$, $\varepsilon$ is the emissivity of the surface and is generally assumed to be 0.95 for our samples (consistent with our temeprature measurements of the disks using a thermal infrared camera), and $T_\infty$ is the ambient temperature.

Defining conduction heat transfer with one general formula for all pressure ranges requires combining free molecular and continuum regimes. The combined form is presented as[4]

$$Q_{cond} = \frac{1}{1/Q_{ccond,co} + 1/Q_{cond,fm}}. \tag{4}$$

Within the continuum[5] and free molecular regime[6], the conduction heat transfer for a disk of radius $a$ is given by

$$Q_{cond,co} = 8ak_{air}(T_s - T_\infty) \tag{5}$$

$$Q_{cond,fm} = Sh_{mol}(T_s - T_\infty), \tag{6}$$

respectively. In Eq. (6), we use the molecular heat transfer coefficient $h_{mol} = \frac{\bar{\alpha}}{8}\frac{\gamma+1}{\gamma-1}\frac{P\bar{v}}{T}$ with the average thermal accommodation coefficient of the top and bottom sides of the disk $\bar{\alpha} = \frac{\alpha_{top}+\alpha_{bottom}}{2}$, the adiabatic constant $\gamma = \frac{C_P}{C_V} = \frac{C_P}{C_P-R}$, and the average speed of gas molecules $\bar{v} = \sqrt{\frac{8k_BT}{\pi m}} = \sqrt{\frac{8R_{air}T}{\pi}}$, where $k_B$ and $m$ are Boltzmann constant and molecular mass of the gas molecules. In calculating $C_P, C_V, k_{air}$ and $\bar{v}$, we approximate the temperature as the average temperature between ambient temperature $T_\infty$ and the surface temperature $T_s$.

In addition to conduction, the convective heat transfer can be written as

$$Q_{conv} = Nu\, a\pi k_{air}(T_s - T_\infty), \tag{7}$$

with $Nu = 0.417Ra^{0.25}$ [7]. The Rayleigh Number is given by $Ra = g\beta(T_s - T_\infty)\frac{\rho d^3}{\mu D}$, where $g$ is gravitational acceleration, $d$ is the diameter of the disk, $\rho$ is density, and $\alpha, \beta,$ and $\mu$ are defined in (1c). As a result, this convection term scales with pressure as $Q_{conv} \propto Ra^{0.25} \propto (P^2)^{0.25} \propto P^{0.5}$ and vanishes in the free molecular regime ($P \to 0$).

Inserting the three heat transfer mode Equations (3), (4), and (7) into Equation (2), we can find the temperature of the surface of a disk, $T_s$, numerically as a function of radius, pressure, and incident intensity.

*Force Formulation:*

The photophoretic force acting on a disk with a temperature difference between the top and bottom sides, which we can call $\Delta T$-force, has been studied extensively[8]. Modifying the surface to achieve different accommodation coefficients on the top and bottom can result in a force on the same order of magnitude, which we will call $\Delta\alpha$-force. Surface modification for a thin lightweight disk is far simpler than fabricating thicker ultralight structures with low thermal conductivity, such as nanocardboard[9].

In both the free molecular regime and the continuum regime, due to similar physical origin of the photophoretic force, the net force on the structure can be expressed as

$$F = \Delta\theta(P)\psi(P), \tag{8}$$

where $\Delta\theta$ is the temperature variation of gas molecules next to the surface, and $\psi\,[\frac{N}{K}]$ represents the force per unit change in temperature of the colliding molecules. $\Delta\theta(P)$ and $\psi(P)$ are both functions of pressure. Equation (8) is based on the interaction between the disk surface and the gas molecules next to the surface.

*Free Molecular Regime*:

In the free molecular limit with $Kn \to \infty$, the average temperature of the gas molecules next to the surface is approximated as $\theta = \frac{1}{2}(T_i + T_r)$ [1], in which $T_i$ and $T_r$ are temperatures of the gas molecules before and after collision, respectively. We can rewrite $\theta$ using the definition of thermal accommodation coefficient between gas molecules and surface, $\alpha = \frac{T_r - T_i}{T_s - T_i}$, which results in $\theta = T_i + \frac{1}{2}\alpha(T_s - T_i)$. Thus, the temperature variation between the two sides of the disk is

$$\Delta\theta(P) = \frac{1}{2}\Delta\alpha(T_s - T_i), \tag{9}$$

for a disk with an accommodation coefficient difference of $\Delta\alpha = \alpha_{bottom} - \alpha_{top}$. In this limit, collision of gas molecules with surface is far more probable than collision of gas molecules with each other, hence the temperature of gas molecules before colliding with surface can be assumed to be equal to far-field temperature, or $T_i = T_\infty$. The force can be found in the free molecular regime with $Kn \to \infty$ [8,10]. The derivation starts by finding the force due to the momentum transfer between the gas molecules and surface. Assuming a uniform temperature across the thickness of the disk and an accommodation coefficient difference of $\Delta\alpha$, we can integrate the Maxwell distribution, $f(v)$, over the entire range of velocity and assume an area of $\pi a^2$ and volume of $V$ of the air with $N$ number of gas molecules. The net force on one side will become:

$$<F> = \pi a^2 N \int_0^\infty \{\frac{(2mv)f(v)v}{V}\}dv \tag{10}$$

and molecule flux of

$$<J> = \frac{N}{V}\int_0^\infty f(v)v\,dv \tag{11}$$

representing the flux of air molecules hitting and reflecting from the surface. In these relations, $f(v)$ is Maxwell distribution and is defined by:

$$f(v) = (\frac{m}{2\pi kT})^{1/2} e^{\frac{-mv^2}{2kT}}. \tag{12}$$

This approach results in the following net $\Delta\alpha$-force on a thin plate with uniform temperature and different accommodation coefficients on two sides [8,10]:

$$F_{fm} = \frac{\pi a^2}{4T_\infty} P\Delta\alpha(T_s - T_\infty). \tag{13}$$

Equation (13) represents a linear increase with pressure which is valid only if the air molecules do not collide with each other as frequently as they do with the surface ($Kn \gg 1$).

*Continuum Regime*:

Here, we will extend the derivation of the photophoretic force acting on a sphere[1] to the case of an oblate spheroid, and then take the limit to approach a flat disk with negligible thickness. In order to find the force for the entire range of pressure, we go back to Equation (8). Knowing the temperature solution from the heat transfer model, we will first find $\Delta\theta(P)$ and then derive an expression for the force.

Continuum regime, part A: constructing $\Delta\theta(P)$:

Consider an oblate spheroid with semi-axes $a$ and $b$ ($a > b$). Similar to the free molecular regime, the average temperature of the gas molecules next to the surface in the continuum regime is approximated as $\theta = \frac{1}{2}(T_i + T_r) = T_i + \frac{1}{2}\alpha(T_s - T_i)$. In the case of the spheroid, mathematical modeleing would not allow for a discontinuity in the value of accommodation coefficient. Thus, in order to achieve a smooth transition from one value of accomodation coefficient to the other instead of the two constant accommodation coefficient on the two sides, the variation of the accommodation coefficient over the surface of the spheroid is approximated by the Legendre expantion, , or:

$$\alpha = \sum a_n P_n(\cos\eta) = a_0 + a_1 \cos\eta + \cdots \tag{14}$$

Finding the coefficient of the Legendre expantion gives $a_0 = \frac{\alpha_{top} + \alpha_{bottom}}{2}$ and $a_1 = \frac{3}{4}\Delta\alpha$ [1]. Finally, the amplitude of temperature variation along the surface, $\Delta\theta$, of the spheroid can be expressed as:

$$\Delta\theta = \frac{3}{8}\Delta\alpha\,(T_s - T_i). \tag{15}$$

In order to express $(T_s - T_i)$ in terms of $(T_s - T_\infty)$, we first construct a form for the conductive heat transfer from the disk as a function of $(T_s - T_\infty)$, then we equate that to the amount of heat being removed from the surface by interaction of the surface with the gas molecules with $T_i$ as their initial temperature right before colliding with the surface. This heat transfer is expressed by [11,12]:

$$\frac{dQ}{dS} = h_{mol}(T_s - T_i), \tag{16}$$

with $dS$ being the surface area element, and $h_{mol}$ molecular heat transfer coefficient defined above. Note that equation (16) holds for the entire range of pressure because the conduction from the surface to the adjacent gas molecules directly on the surface happens *via* molecular interaction. Once $Q$ is found as a function of $(T_s - T_\infty)$, equation (16) can be used to find $(T_s - T_i)$ which gives $\Delta\theta$ according to equation (15).

In the continuum regime, $Q_{cond,co}$ does not have a trivial solution for a spheroid. We set up the problem of steady heat conduction around a spheroid of surface temperature $T_s + \Delta\theta\cos(\eta)$, which is a superposition of a uniform value and a surface varying component with $\eta$ as the polar angle of the spheroid, in a medium of ambient temperature $T_\infty$ and without volumetric heat generation within the spheroid. The governing is

$$\frac{1}{\cosh\xi}\frac{\partial}{\partial\xi}\left[\cosh\xi\,\frac{\partial T}{\partial\xi}\right] + \frac{1}{\sin\eta}\frac{\partial}{\partial\eta}\left[\sin\eta\,\frac{\partial T}{\partial\eta}\right] = 0, \tag{17}$$

where $\xi$ and $\eta$ are radial and angular parts of spheroidal coordinates[13] (see Figure S5). The boundary conditions for this problem are

$$T = T_s + \Delta\theta \cos(\eta) \quad \text{at} \quad \xi = \xi_0, \quad \text{(surface of the spheroid, } \xi_0 = \tanh^{-1}\frac{b}{a}\text{)}$$

$$T = T_\infty \quad \text{as} \quad \xi \to \infty.$$
(18)

The temperature solution for equation (17) with boundary conditions shown in (18) becomes:

$$T(\xi,\eta) - T_\infty = (T_s - T_\infty)\frac{\tan^{-1}(\sinh\xi) - \frac{\pi}{2}}{\tan^{-1}(\sinh\xi_0) - \frac{\pi}{2}} + \Delta\theta \cos(\eta)\frac{\sinh(\xi)(\tan^{-1}(\sinh\xi) - \frac{\pi}{2}) + 1}{\sinh(\xi_0)(\tan^{-1}(\sinh\xi_0) - \frac{\pi}{2}) + 1}.$$
(19)

The local amount of heat transfer can be found using solution (19) and its proper boundary conditions:

$$[\nabla T]_{\xi=\xi_0} = \frac{1}{l}\left[\frac{\partial T}{\partial \xi}\hat{e}_\xi + \frac{\partial T}{\partial \eta}\hat{e}_\eta\right]_{\xi=\xi_0} = \frac{1}{l}\left[(T_s - T_\infty)\frac{1}{\cosh\xi_0\{\tan^{-1}(\sinh\xi_0) - \frac{\pi}{2}\}}\hat{e}_\xi - \Delta\theta \sin(\eta)\,\hat{e}_\eta\right]$$
(20)

with $l = \sqrt{(a^2 - b^2)(\sinh^2\xi_0 + \cos^2\eta)}$. It should be noted that the normal component of the temperature gradient due to surface-varying component of the temperature is not included in (20) since its integration over the surface of the spheroid is zero. Thus, the total heat flow from the surface of the oblate spheroid

$$Q_{cond,co} = \int(-k_{air}\nabla T \cdot \hat{e}_\xi)dS = \frac{8\pi a k_{air}(T_s - T_\infty)}{sh},$$
(21)

where, $sh$ is a shape factor that depends solely on geometry of the sample and defined as

$$sh = \frac{(\pi - 2\tan^{-1}(\sinh\xi_0))}{\cosh\xi_0}.$$
(22a)

In the two limiting cases of disk and sphere, the shape factor reduces to

$$\lim_{\frac{b}{a}\to 0}(sh) = \pi \quad \text{for a disk.} \tag{22b}$$

$$\lim_{\frac{b}{a}\to 1}(sh) = 2 \quad \text{for a sphere.} \tag{22c}$$

In the case of a disk, equation (21) reduces to Eq. (5): $Q_{cond,co} = 8ak_{air}(T_s - T_\infty)$.

Now, relating this heat conduction from the disk to the total heat transfer from (16), we find,

$$h_{mol}(T_s - T_i)2\pi a^2 = 8ak_{air}(T_s - T_\infty), \tag{23}$$

thus, the temperature difference in (15) reduces to

$$\Delta\theta_{disk} = \frac{3}{2\pi}\frac{k_{air}}{ah_{mol}}\Delta\alpha(T_s - T_\infty). \tag{24}$$

*Continuum regime, part B: developing the force formula, $F = \Delta\theta(P)\psi(P)$:*

The photophoretic force on a particle in continuum regime is caused by thermal creep[14,15]. When the gas over a surface has a tangential temperature gradient, it flows over the surface from the cooler side to the hotter side with slip velocity, $\mathbf{v}_s$, defined by:

$$\mathbf{v}_s = \kappa_s \frac{\mu_{air}}{\rho T_\infty} \nabla_s T, \quad (25)$$

where $\kappa_s = 1.14$ is thermal slip coefficient, $\mu$ is viscosity, $\rho$ is density, and $\nabla_s T$ is the tangential temperature gradient in the gas layer. Using (20), we have

$$\nabla_s T = -\frac{\Delta\theta}{\sqrt{(a^2-b^2)(\sinh^2\xi_0+\cos^2\eta)}}\sin(\eta)\,\hat{\mathbf{e}}_\eta. \quad (26)$$

In order to calculate the force, we use the Lorentz reciprocal theorem[14] for the Stokes flow[15] we can find the migration velocity of the spheroidal particle along its symmetry axis by

$$U = -\frac{1}{4\pi b a^2}\int (\mathbf{n}\cdot\mathbf{r})(\mathbf{v}_s\cdot\hat{\mathbf{e}}_z)\,dS = \frac{\kappa_s\mu}{\rho T_\infty}\frac{\Delta\theta}{a}\sinh\xi_0\cosh\xi_0\left(\frac{\cosh^2\xi_0}{\sinh\xi_0}\tan^{-1}\left(\frac{1}{\sinh\xi_0}\right)-1\right). \quad (27)$$

By taking the limit of $\frac{b}{a}\to 0$, we find the value of migration velocity for a disk, to be:

$$U_{disk} = \frac{\pi\kappa_s\mu_{air}}{2a\rho T_\infty}\Delta\theta = \frac{\pi\kappa_s\mu_{air}}{2a\rho T_\infty}\frac{3}{2\pi}\frac{k_{air}}{a\frac{\bar{\alpha}\gamma+1 P\bar{v}}{8\gamma-1 T}}\Delta\alpha(T_s-T_\infty) = \frac{6\kappa_s\mu T_s k_{air}}{a^2\rho\, P\bar{v}}\frac{\gamma-1}{\gamma+1}\frac{\Delta\alpha}{\bar{\alpha}}\frac{(T_s-T_\infty)}{T_\infty}. \quad (28)$$

Using kinetic theory of gases[16] we can substitute $k_{air} = fC_v\mu_{air}$ and $\mu_{air} = \frac{1}{2}\bar{v}\rho\lambda$, where we use the standard definition for the mean free path $\lambda = \frac{\mu_{air}}{P}\sqrt{\frac{\pi R_{air}T}{2}}$ [17], the $f$-factor is given by $f = 1 + \frac{9R_{air}}{4C_v} = 1 + \frac{9}{4}(\gamma-1)$ [16], and we substituted $\frac{R}{C_v} = \frac{C_p-C_v}{C_v} = \frac{C_p}{C_v}-1 = \gamma-1$. The resulting expression is

$$U_{disk} = \bar{v}\frac{\kappa_s f \rho T_s C_v}{P}\left(\frac{\lambda}{a}\right)^2\frac{\gamma-1}{\gamma+1}\frac{\Delta\alpha}{\bar{\alpha}}\frac{T_s-T_\infty}{T_\infty} = \bar{v}\kappa_s f\frac{C_v}{R_{air}}\left(\frac{\lambda}{a}\right)^2\frac{\gamma-1}{\gamma+1}\frac{\Delta\alpha}{\bar{\alpha}}\frac{T_s-T_\infty}{T_\infty} = \bar{v}\kappa_s f\left(\frac{\lambda}{a}\right)^2\frac{1}{\gamma+1}\frac{\Delta\alpha}{\bar{\alpha}}\frac{T_s-T_\infty}{T_\infty} = $$
$$\bar{v}\,\kappa_s\left(\frac{\lambda}{a}\right)^2\frac{1+\frac{9}{4}(\gamma-1)}{\gamma+1}\frac{\Delta\alpha}{\bar{\alpha}}\frac{T_s-T_\infty}{T_\infty}, \quad (29)$$

where we again used the ideal gas law, $P = \rho T_s R_{air}$. The last expression in Eq. 29 suggest that in the transition regime ($\frac{\lambda}{a}\sim 1$), the air can flow around the disk at a significant fraction of the average speed of the air molecules (i.e., tens of meters per second).

Equation (28) represents velocity of a disk that is free to move in gaseous medium without any forces acting on it. For a fixed (immobile) disk, the corresponding force acting on the disk can be obtained using Stokes drag formula with effective Stokes radius for a disk[18], $r = \frac{8}{3\pi}a$:

$$F_{co} = 6\pi\mu_{air}\left(\frac{8}{3\pi}a\right)U_{disk} = 16\mu_{air}aU_{disk} = 16\mu\frac{6\kappa_s\mu_{air}T_s k_{air}}{a\rho\,P\bar{v}}\frac{\gamma-1}{\gamma+1}\frac{\Delta\alpha}{\bar{\alpha}}\frac{(T_s-T_\infty)}{T_\infty}$$
(30)

As a check to see whether the Stokes flow assumption is correct, we can evaluate the Reynolds Number using the migration velocity in (28), $Re = \frac{\rho U_{disk} 2a}{\mu} = \pi\kappa_s \frac{\Delta\theta}{T_\infty} \ll 1$, which justifies the Stokes Flow assumption.

*Force formula for the entire range of pressure:*

Finally, we follow the example of Loesche et al[19] and combine the photophoretic force in the free molecular (13) and the continuum (30) regimes to generate an interpolation valid for the entire range of pressure,

$$F = \frac{1}{\frac{1}{F_{fm}}+\frac{1}{F_{co}}} = \frac{\pi}{4}\frac{\Delta\alpha(T_s-T_\infty)}{T_\infty}\frac{a^2 P^*}{\frac{P^2}{2P^{*2}}+\frac{P^*}{P}}$$
(31)

where $P^*$ is the pressure at which force is maximized:

$$P^* = \frac{1}{a}\left(\frac{192\,\mu_{air}^2 k_B(\gamma-1)k_{air}T_{avg}^2}{\pi\,m(\bar{\alpha})(\gamma+1)\bar{v}}\right)^{\frac{1}{3}} = \frac{\bar{v}\rho\lambda}{2a}\left(\frac{48(\gamma-1)(9\gamma-5)}{\pi m\bar{\alpha}\bar{v}}C_v k_B T_{avg}^2\right)^{\frac{1}{3}}.$$
(32)

All the parameters are the same as defined earlier in the text. It is worth noting that $P^* \propto \frac{1}{a}$ and, therefore, the maximum force

$$F_{max} = \frac{\pi}{6}\frac{\Delta\alpha(T_s-T_\infty)}{T_\infty}a^2 P^*$$
(33)

scales linearly with the radius of the disk if all other parameters are held constant.

*Altitude dependency of the properties for the Earth's atmosphere:*

In order to model changes in temperature and pressure of ambient air as a function of altitude, we incorporated altitude dependency in all parts of the model that are functions of temperature and pressure. Figure S6 shows how the temperature and pressure depend on the altitude based on[20]. We note that these graphs represent annual and spatial averages and the values may vary depending on the exact location and time of year.

*Predicted payload for various combinations of emissivity and $\Delta\alpha$:*

In addition to the predicted payload shown in Fig. 4 of the main text, we also calculated the maximum payload for other combinations of emissivity and the difference in the thermal accommodation coefficient. Fig. S8 shows the predicted payload for the parameters that provided the best fit to our actual experiments (with $\Delta\alpha = 0.15$, $\epsilon = 0.95$). In this case, the maximum payload is comparable to the weight of the disk itself (~0.1 mg) and is achieved for disk radius of ~1 cm at altitudes of ~80 km. Figure S9 shows that reducing the thermal emissivity using a selective solar absorber with $\epsilon = 0.5$ increases the maximum payload to ~0.5 mg, still achieved for a radius of ~1 cm at altitudes of ~80 km. Figure S10 shows that using an even lower emissivity of 0.1 makes the temepratures exceed 500 K, which would likely require the use of materials other than Mylar. However, such low emissivities also allow levitation and significant payloads with much lower ligh intensities that full natural sunlight (Fig. S11).

Increasing the Δ$\alpha$ to 0.3 improves the maximum payload to a few mg, achieved for radii of a few cm at altitudes of ~85km (Fig. S12 and S13). Finally, using Δ$\alpha$=0.5 results in maximum payloads of up to 10 mg for radii of ~3cm at alututudes of ~90 km (Fig. S14 and Figure 4 of the main text).

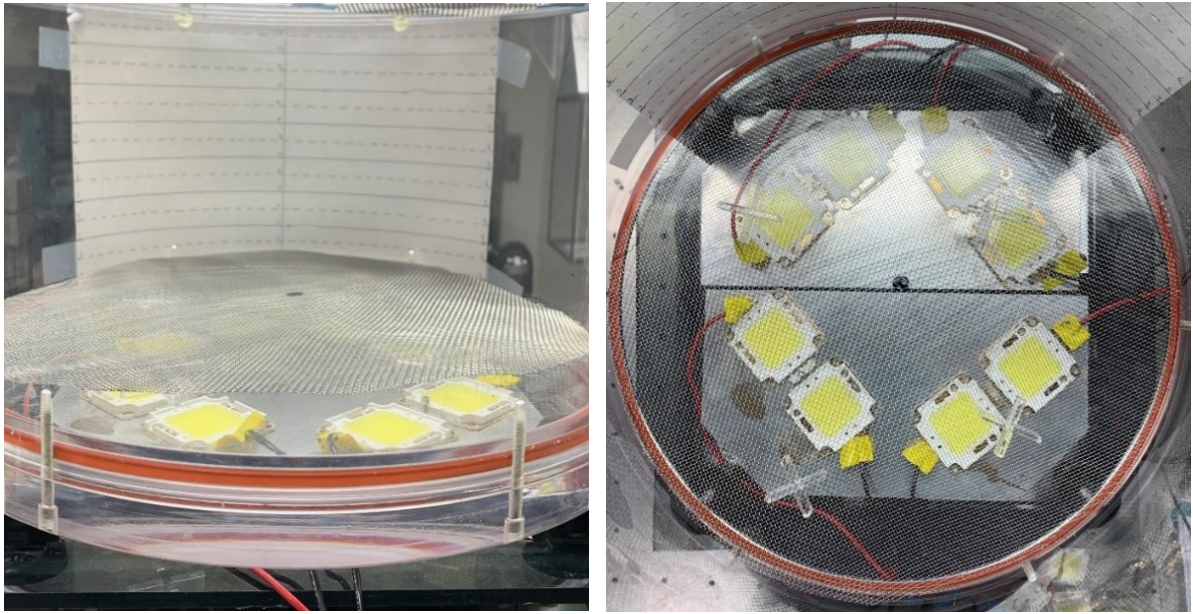

**Figure S1. left)** Side view of the chamber and sample. **Right)** Top view of the setup with the 8 array of LED light ring and 74% transparent mesh.

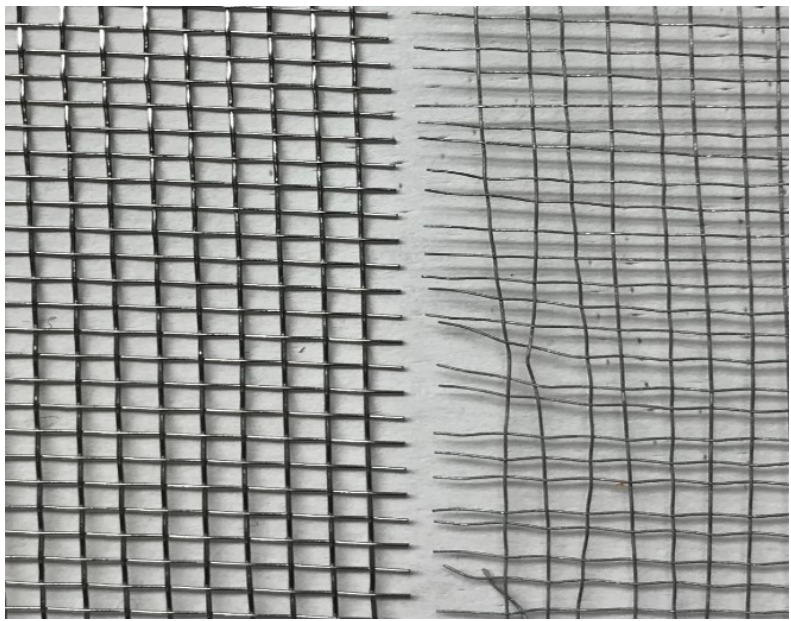

**Figure S2**. Metal meshes that were used as launchpads: The left one has 74% open area while the right one is 85% open.

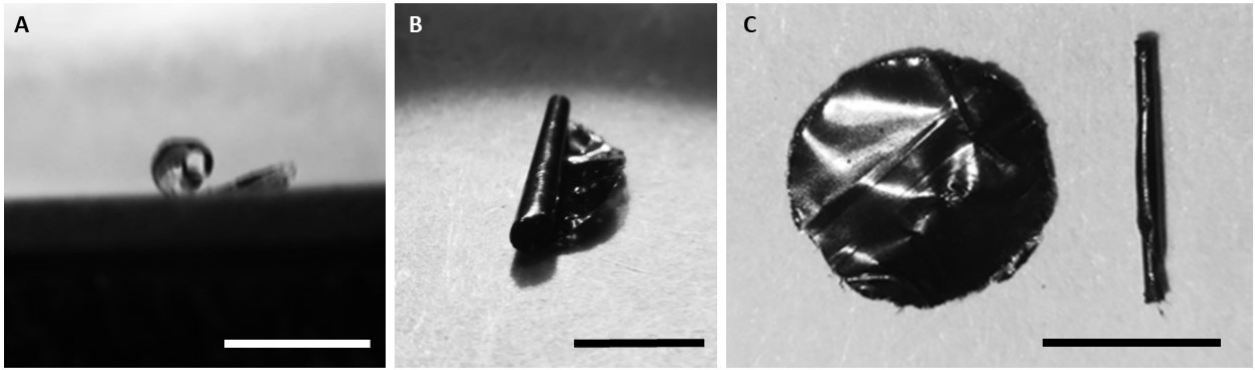

**Figure S3.** Thermal deformation of mylar samples at high intensities. **A)** and **B)** 6 mm disk under $6\ \frac{kW}{m^2}$ in $50\ Pa$ environment. **C)** left) undeformed and right) deformed 6 mm disk. The highly curled sample, which has rolled up into a cylinder with submillimeter diameter has been under $8\ \frac{kW}{m^2}$ in $50\ Pa$ environment. All scale bars are 3 mm.

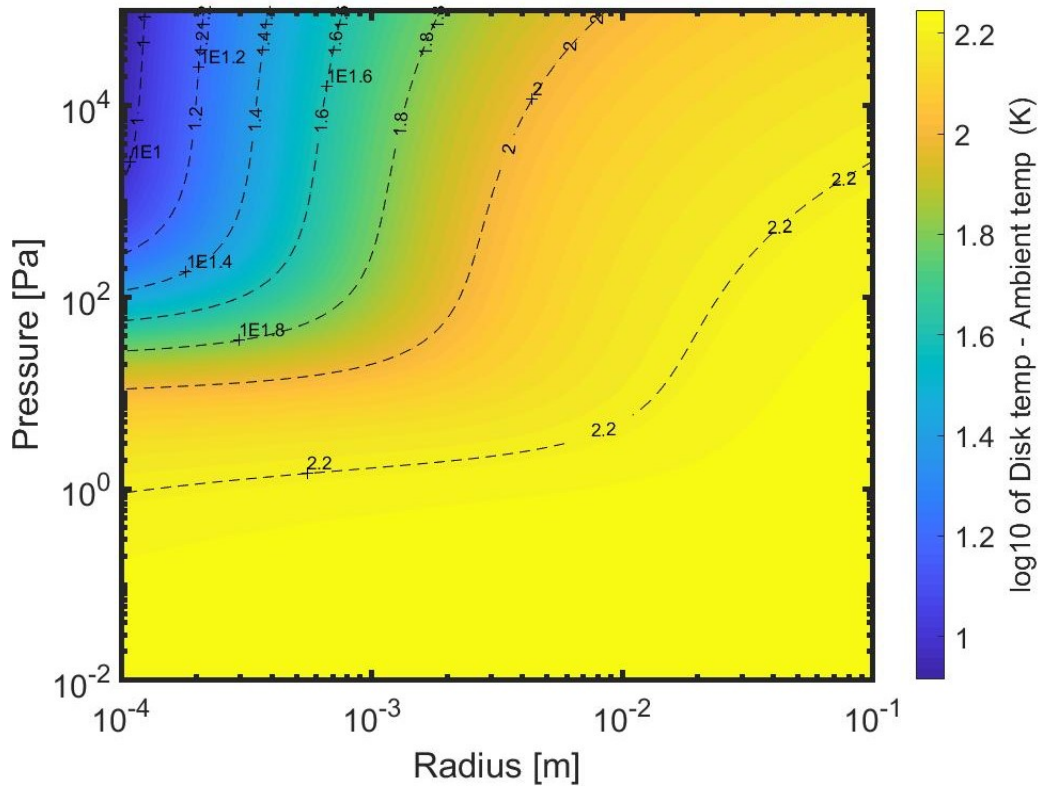

**Figure S4.** Calculated temperature of mylar disk under $5\ \frac{kW}{m^2}$ incident light with $\epsilon = 0.95$. This plot was used to predict at what pressures, and with what radii, samples exceed a temperature threshold and deform.

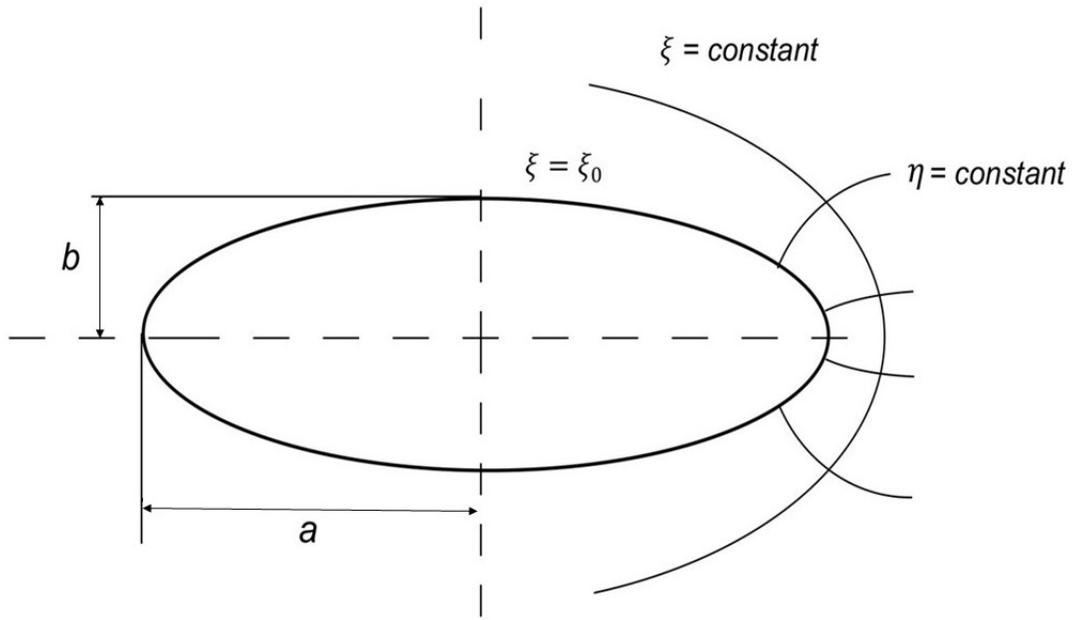

**Figure S5**. Spheroidal coordinate used in theory development.

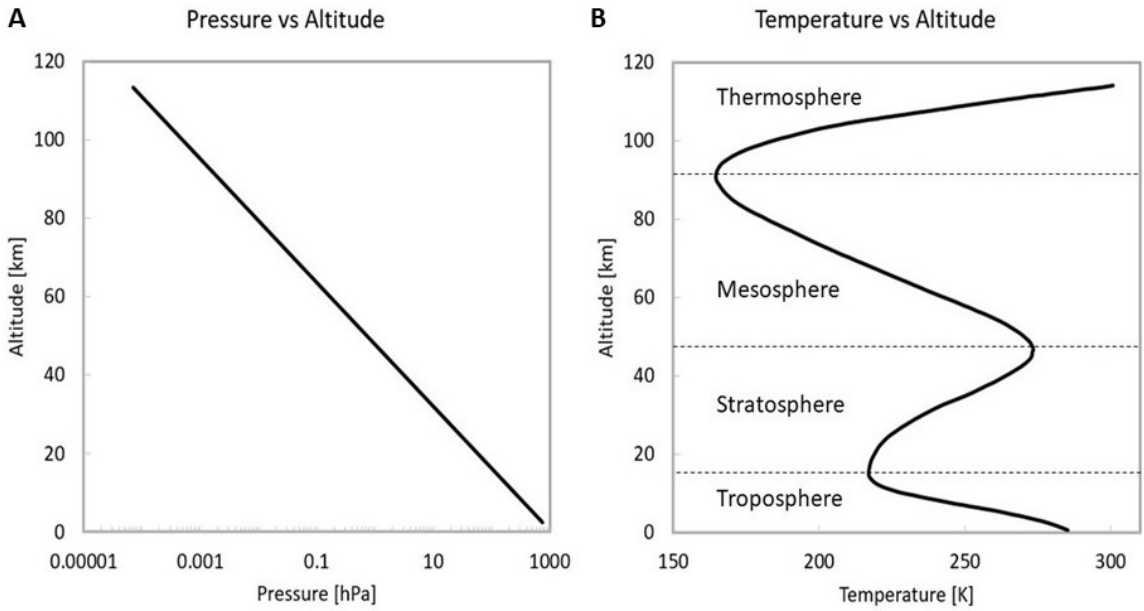

**Figure S6.** Altitude dependence **A)** Pressure and **B)** Temperature used in the calculation[20].

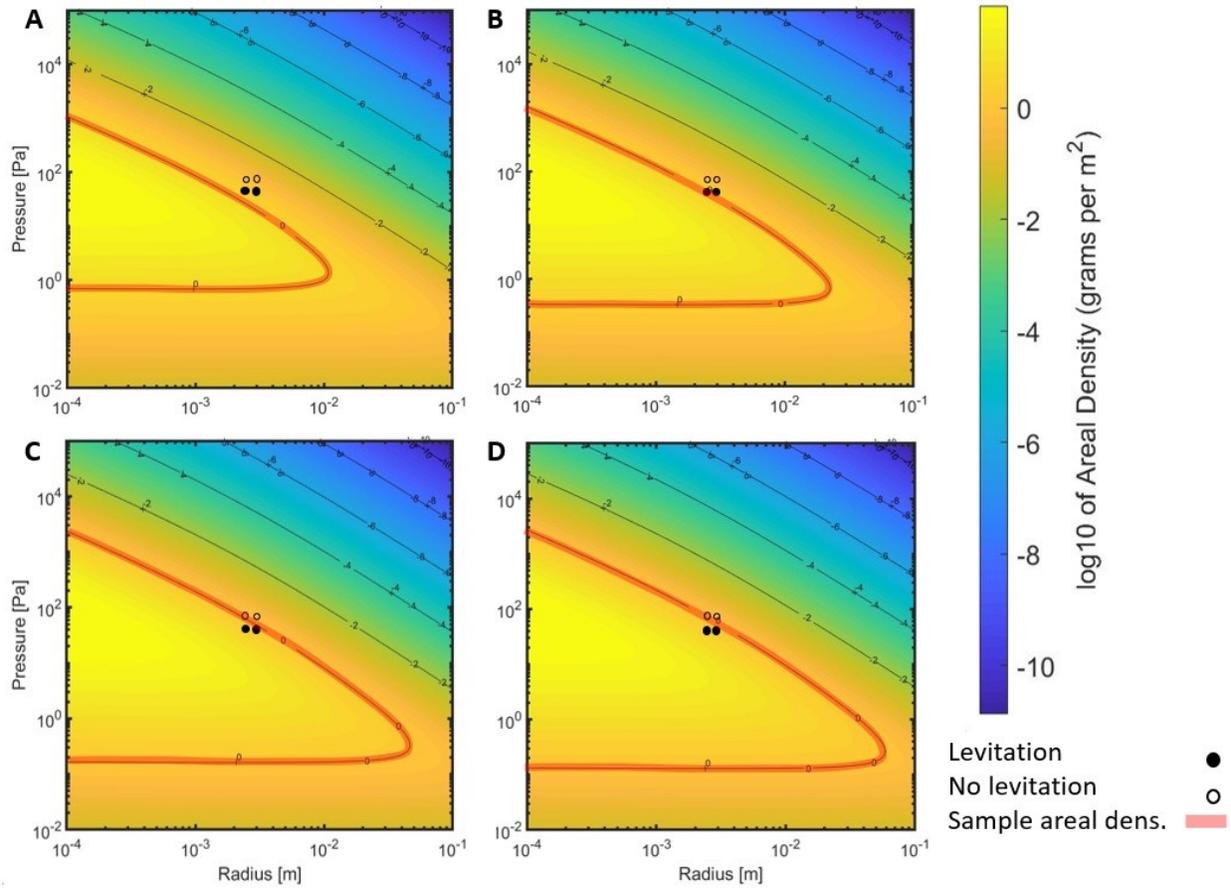

**Figure S7**. Areal density that can be levitated under $0.5 \frac{W}{cm^2}$, with $\epsilon = 0.95$ and **A)** $\Delta\alpha = 0.05$ **B)** $\Delta\alpha = 0.1$ **C)** $\Delta\alpha = 0.2$ **D)** $\Delta\alpha = 0.25$. The range of $0.1 < \Delta\alpha < 0.2$ results in an acceptable match between the experimental observations and theoretical predictions.

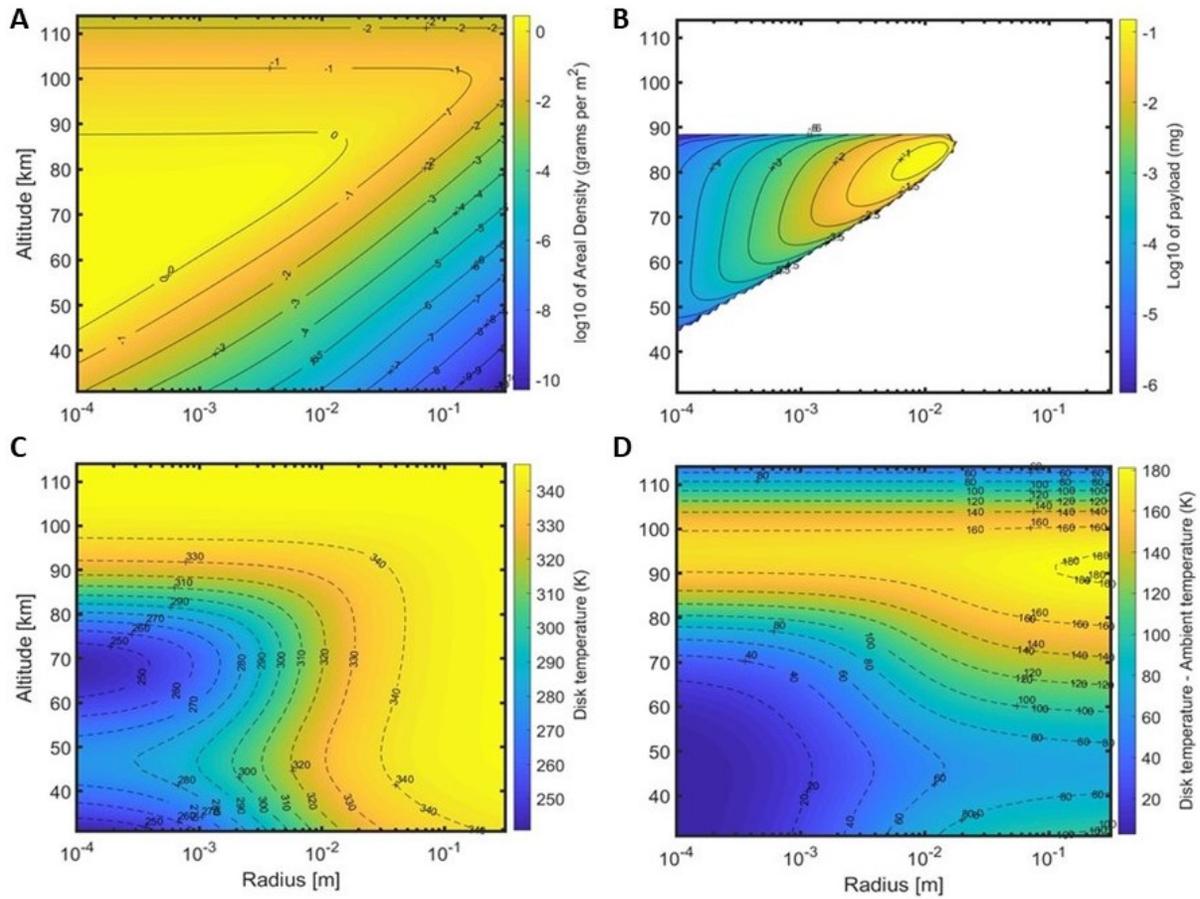

**Figure S8.** Contour plots of **A)** Areal density of the object able to be levitated **B)** Payload that can be lifted using mylar-CNT **C)** Temperature and **D)** Temperature difference between the disk and ambient for different sizes at different altitudes with $\Delta\alpha = 0.15$, $\epsilon = 0.95$ and under natural sunlight ($0.136 \frac{W}{cm^2}$).

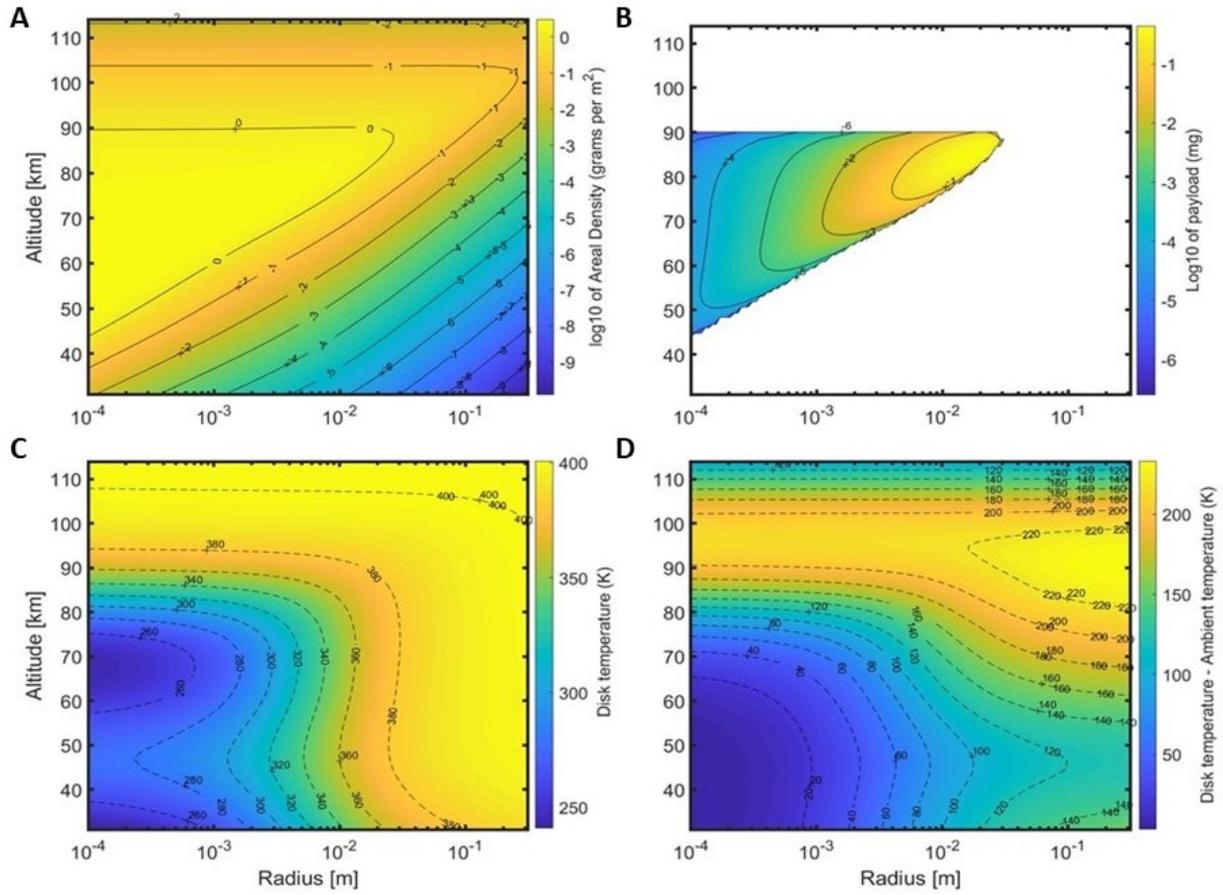

**Figure S9.** Same as Figure S8 for $\Delta\alpha = 0.15$, $\epsilon = 0.5$ and under natural sunlight ($0.136 \frac{W}{cm^2}$).

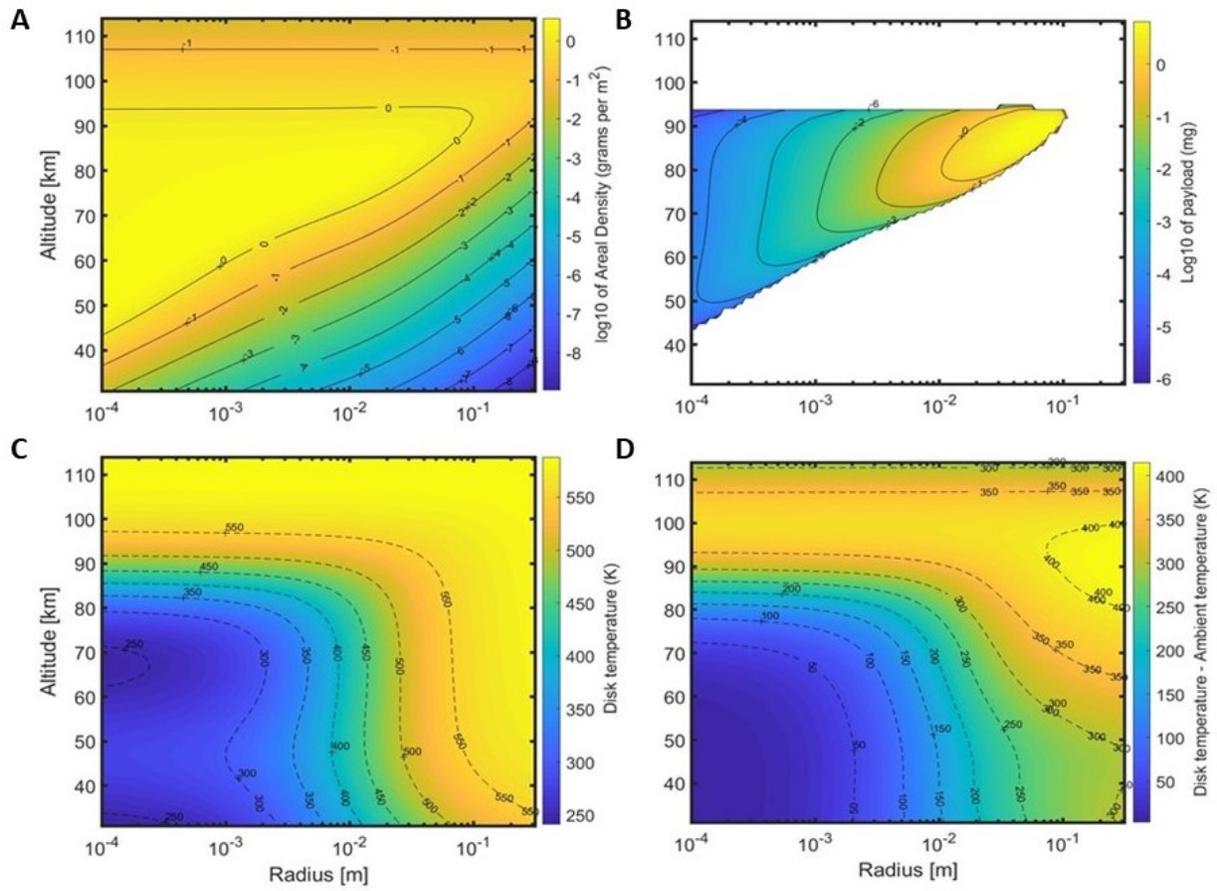

**Figure S10.** Same as Figure S8 for $\Delta\alpha = 0.15$, $\epsilon = 0.1$ and under natural sunlight $(0.136\frac{W}{cm^2})$.

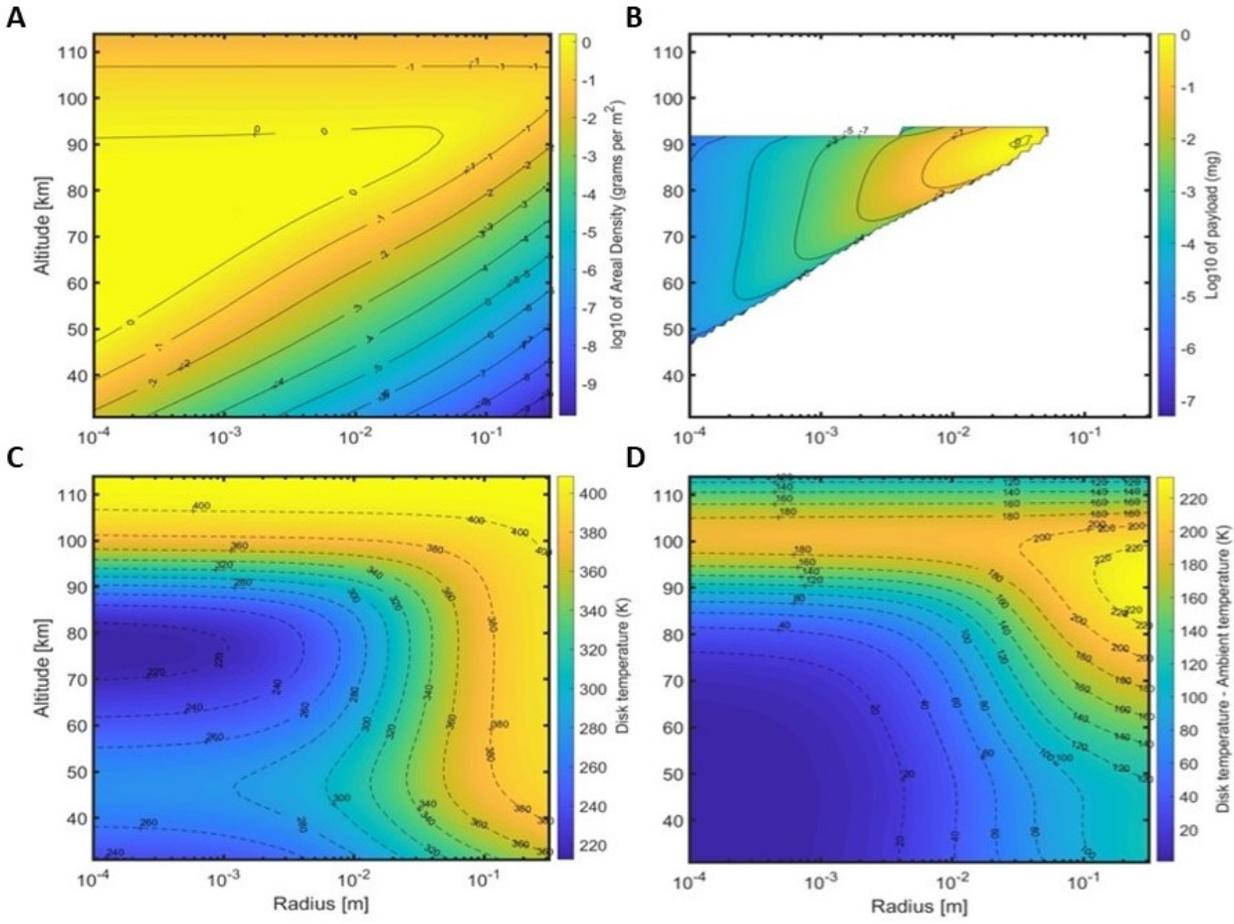

**Figure S11.** Same as Figure S8 for $\Delta\alpha = 0.3$, $\epsilon = 0.1$ and under light intensity about 4.5 times less than natural sunlight ($0.03 \frac{W}{cm^2}$).

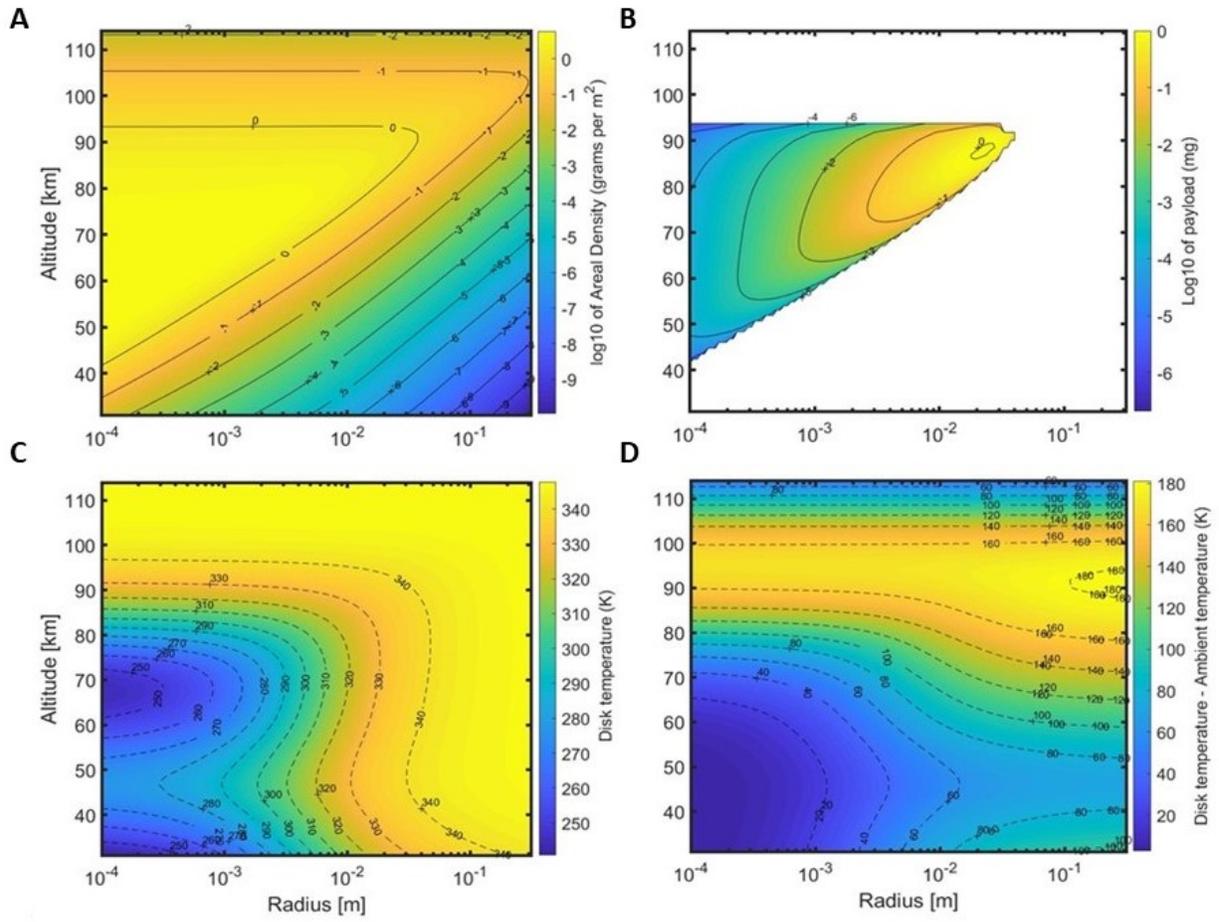

**Figure S12.** Same as Figure S8 for $\Delta\alpha = 0.3$, $\epsilon = 0.95$ and under natural sunlight ($0.136 \frac{W}{cm^2}$).

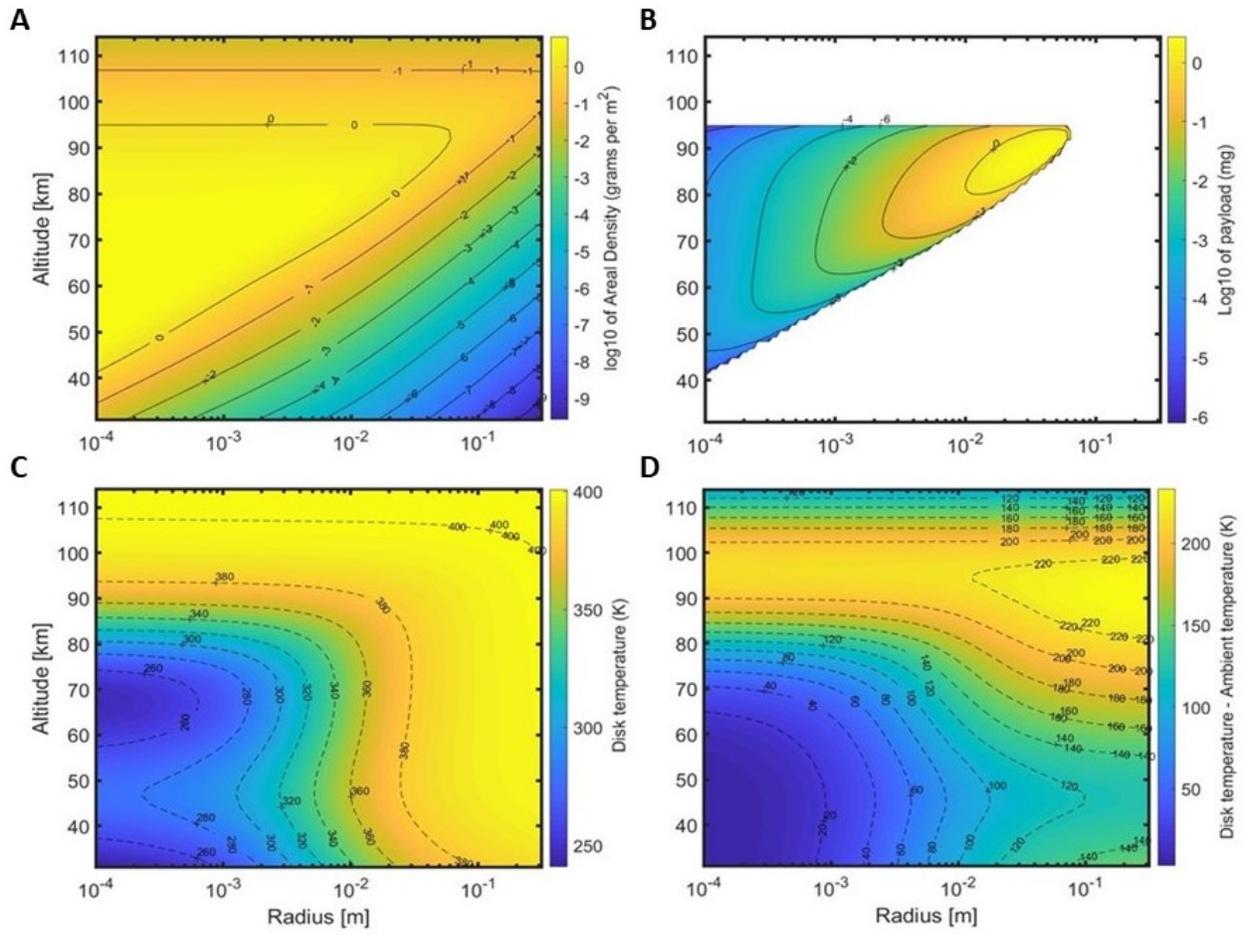

**Figure S13.** Same as Figure S8 for $\Delta\alpha = 0.3$, $\epsilon = 0.5$ and under natural sunlight ($0.136 \frac{W}{cm^2}$).

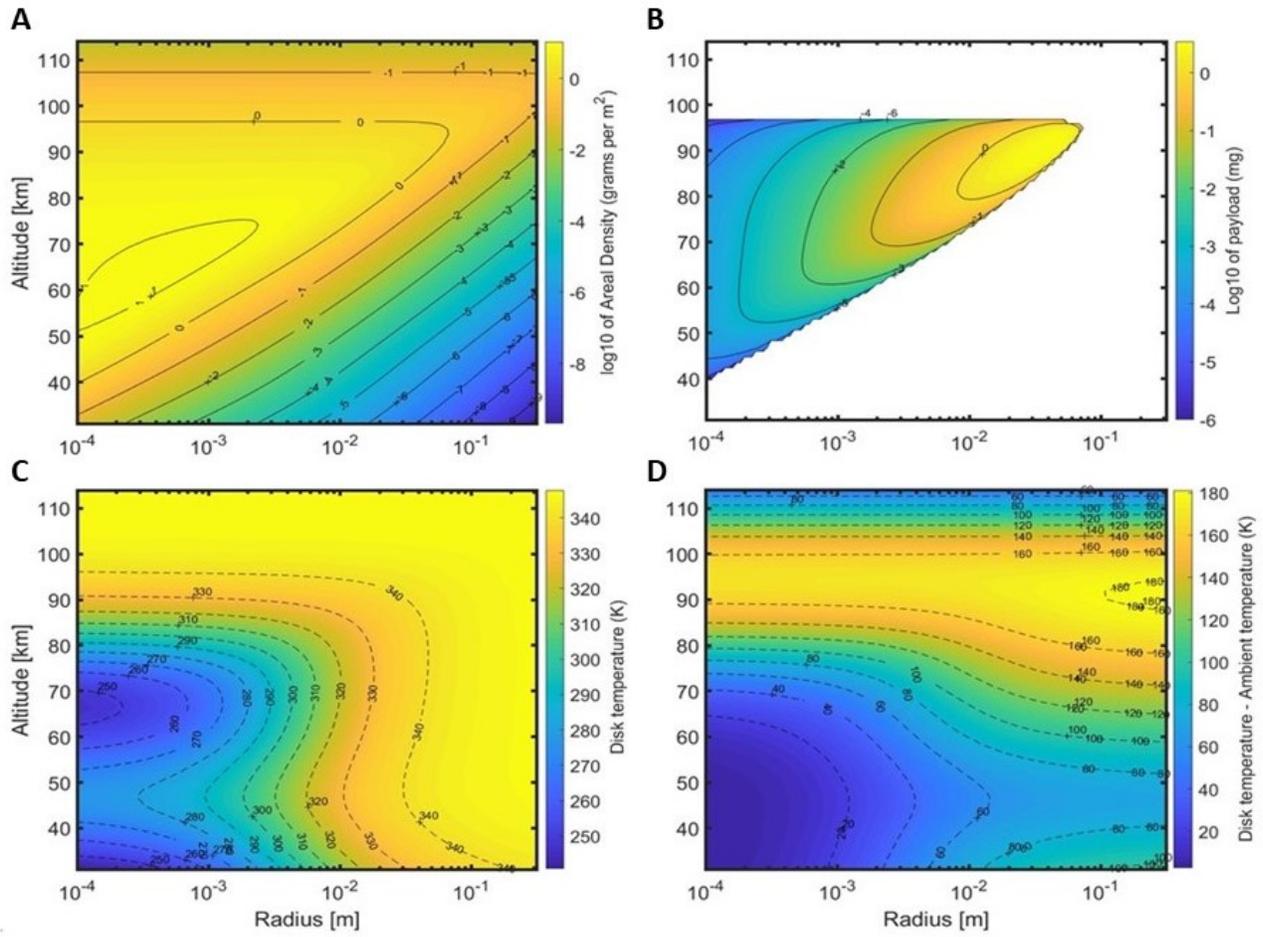

**Figure S14.** Same as Figure S8 for $\Delta\alpha = 0.5$, $\epsilon = 0.95$ and under natural sunlight ($0.136\frac{W}{cm^2}$).

**Movie S1.**

Movie S1 presents a side view of a 6-mm-diameter disk levitated at 20 Pa under a light intensity of $0.5 \frac{W}{cm^2}$ (~ 5 suns). The test was done with the small light trap (4-cm-diameter light ring), which was insufficient to trap the sample and the sample flew out of the trap. The movie is slowed down to $1/8^{th}$ speed.

**Movie S2.**

Movie S2 presents oblique view of a 6-mm-diameter disk levitated at ~ 20 Pa under under a light intensity of $0.5 \frac{W}{cm^2}$ (~ 5 suns) with the large light trap (15-cm-diameter light ring). After several seconds of successful levitation, the intensity was increased gradually (~ 3 $\frac{kW}{m^2}$ per min) to over 6 suns resulting in a temperature increase that slowly deformed the sample. This deformation decreased the effective light absorbing area causing the sample to settle inside the light trap area. From the onset of levitation to landing of the sample due to thermal deformation, the total flight duration was around 30 seconds. The movie is slowed down to $1/8^{th}$ speed.

**Movie S3.**

Movie S3 presents oblique view of a 6-mm-diameter disk levitated at ~ 30 Pa under a light intensity of $0.5 \frac{W}{cm^2}$ (~ 5 suns) with the large light trap. Same as Movie S2, after several seconds of sustained levitation, the intensity was increased gradually to over 6 suns resulting in a temperature increase that slowly deformed the sample. This deformation decreased the effective light absorbing area causing the sample to settle inside the light trap area. From the onset of levitation to landing of the sample due to thermal deformation, the total flight duration was around 30 seconds.

**Movie S4.**

Movie S4 presents side view of two 6-mm-diameter disk levitated at ~ 30 Pa under a light intensity of $0.5 \frac{W}{cm^2}$ (~ 5 suns) with the large light trap. Same as Movie S2, after several seconds of sustained levitation, the intensity was increased gradually to over 6 suns resulting in a temperature increase that slowly deformed the sample. This deformation decreased the effective light absorbing area causing the sample to settle inside the light trap area. From the onset of levitation to landing of the sample due to thermal deformation, the total flight duration was around 30 seconds.

**Movie S5.**

Movie S5 presents side view of a 6-mm-diameter disk at 6 Pa. In this test, we gradually increased the light intensity. Even though the force is predicted to be enough to levitate the sample, the temperature of the disk increased to higher than thermal deformation threshold resulting in deformation of the sample before levitation.